\documentclass[12pt]{article}
\usepackage{amsmath}
\usepackage{graphicx,psfrag,epsf}
\usepackage{enumerate}
\usepackage{natbib}
\usepackage{url} 

\newcommand{\blind}{0}

\usepackage[table,dvipsnames]{xcolor}
\usepackage{amsthm}
\usepackage{mathtools}
\usepackage{collcell}
\usepackage{dsfont}
\usepackage{rotating}
\usepackage{wrapfig}
\usepackage[utf8]{inputenc}
\usepackage{tikz}
\usepackage{lettrine}
\usepackage{dblfloatfix}
\usepackage[english]{babel} 
\usepackage{amsfonts}
\usepackage{subfig}
\graphicspath{ {fig/} }
\usepackage[hyperfootnotes=false]{hyperref}

\usepackage{lipsum}
\usetikzlibrary{arrows,decorations.markings}
\usepackage{standalone}

\begin{document}

	\def\spacingset#1{\renewcommand{\baselinestretch}%
		{#1}\small\normalsize} \spacingset{1}

	
	\if0\blind
	{
		\title{\bf Changes in electricity demand pattern in Europe due to COVID-19 shutdowns}
		\author{Michał Narajewski\hspace{.2cm}\\
			University of Duisburg-Essen\\
			and \\
			Florian Ziel \\
			University of Duisburg-Essen}
		\maketitle
	} \fi
	
	\if1\blind
	{
		\bigskip
		\bigskip
		\bigskip
		\begin{center}
			{\LARGE\bf Title}
		\end{center}
		\medskip
	} \fi
	
	\bigskip
\begin{abstract}
	The article covers electricity demand shift effects due to COVID-19 shutdowns in various European countries. We utilize high-dimensional regression techniques to exploit the structural breaks in demand profiles due to the shutdowns. We discuss the findings with respect to coronavirus pandemic progress and regulatory measures of the considered countries.
\end{abstract}

\noindent%
	\vfill
	
	\newpage
	\spacingset{1.45} 
	
	
	\section{Introduction}
	The ongoing coronavirus pandemic in 2020 and especially the preventive measures to reduce the \mbox{COVID-19} disease changed drastically the patterns of our behaviour. Many countries in Europe and in the world introduced multiple levels of restrictions: companies sent their office employees to work from home, schools and universities closed, many factories limited or stopped their production, curfews and similar stay-at-home orders. All these factors impact the energy demand by decreasing the overall level and changing its behaviour. In this paper, we analyse the change in electricity demand pattern in selected European countries caused by the \mbox{COVID-19} shutdowns.

		\begin{figure}[b!]
	
	\includegraphics[width=1\textwidth]{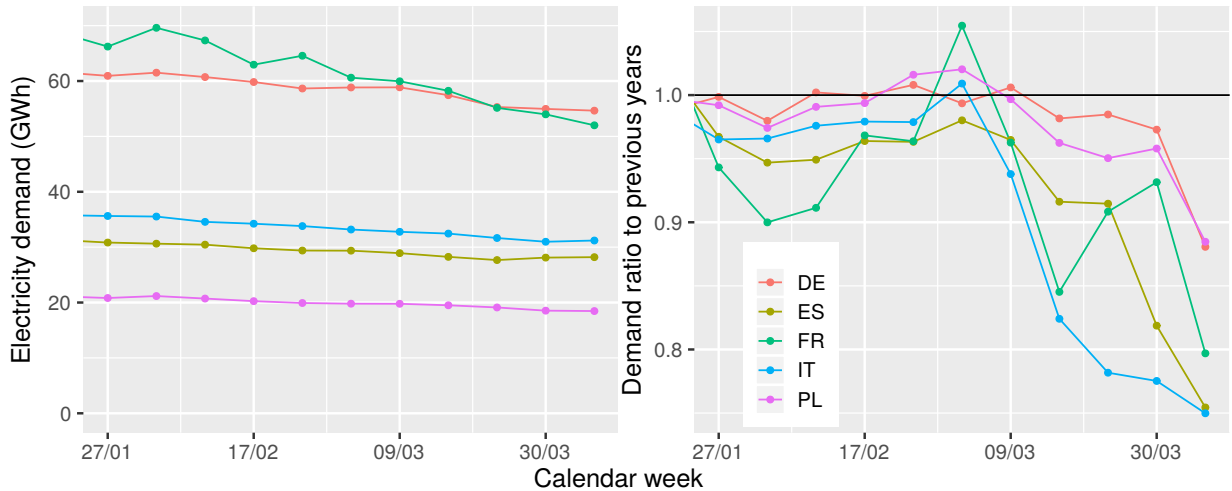}
	
	\caption{Median weekly average electricity demand (GWh) in years 2016-2019 (left) and the 2020 demand ratio to the average in years 2016-2019.}
	\label{fig:motivation}
	
\end{figure}

For the analysis we consider the five most populated countries of the European Union: Germany, France, Italy, Spain and Poland. The spread of the coronavirus as well as the undertaken coronavirus measures are on multiple levels in these countries in spring 2020. Also, the demand shifts are different for each of the countries what is depicted in Figure~\ref{fig:motivation}. In Europe, the pandemic started in Italy and this is also reflected in the electricity load change in Figure~\ref{fig:motivation}. A very high rise of the number of infected people in the beginning of the outbreak resulted in a very strict lock-down in the whole country (\citealp{flaxman2020report}; \citealp{saglietto2020covid}). Thus, we focus particularly on the electricity demand of Italy. The coronavirus started spreading later in the other analysed countries and therefore at the time of this analysis its progress differs significantly -- from a very similar in Spain to much lower in Poland.

	In the next section, we present the data used for the analysis of the electricity demand. Then, the utilized methodology and the model are discussed. The fourth section consists of the results which are presented and analysed separately for Italy and for the other countries. The last section concludes the paper.

	\section{Electricity demand data}
	The data utilized in purpose of this exercise was downloaded from the publicly available \cite{ENTSOEtransparency} Transparency platform. We use the actual total load data of all mentioned countries, and they span the data range from 1 January 2016 to 15 April 2020.\footnote{ Minor missing values were interpolated linearly. Additionally, the data is aggregated in hourly intervals.}
	
	\begin{figure}[b!]
		\vspace{-0.4cm}
		\includegraphics[width=1\linewidth]{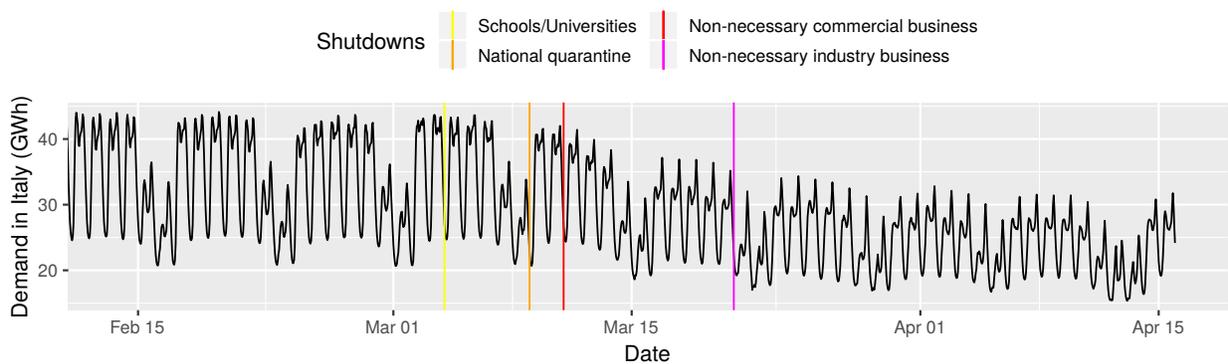}
		
		\caption{Electricity demand (GWh) in Italy during the ongoing pandemic. The vertical lines indicate shutdown dates.}
		\label{fig:italy_timeline}
	\end{figure}
	A small part of the data is presented in Figure~\ref{fig:italy_timeline}. It shows the electricity demand in Italy over time during the ongoing pandemic. Moreover, we highlighted the dates of four nationwide shutdowns. The shutdown of all schools and universities does not seem to have impacted the electricity load in Italy. Only the introduction of the national quarantine and then tightening of the lock-down by closing down all non-essential commercial and retail businesses seem to have first impacted the demand. Then, halting all non-necessary production and industries seem to have deepened the decrease.
	However, the plotted time interval is also the beginning of spring. At this time of the year, a decrease of Italian electricity demand is usually observed. Therefore, in order to recognize whether the change in the load is shutdown-, season-, or weather-driven we need a sophisticated demand model to disentangle the reduction effects.

	\section{Methodology}
	For exploiting the structural changes in the electricity demand due to the shutdown we apply a high-dimensional time series change-point models to the electricity log-load of each country. As baseline for the analysis of the structural changes we consider a model that is very similar to the load forecasting model of \cite{ziel2016lasso} that was successfully applied in the framework of the Global Energy Forecasting Competition 2014 for electricity load forecasting. We refer for technical details to the aforementioned paper. However, here we want to describe the relevant model properties that are important to understand and interpret the results.
	For the analysis we consider a baseline model that assumes no structural changes in the data. Then, this model is augmented by change-point components. First, we describe briefly the baseline model, to proceed with the change-point part.
	
	\subsection{Baseline model}
	The baseline model contains mainly two types of components i) pattern-based time-varying coefficients and ii) autoregressive effects. The time-varying coefficients vary mainly seasonally and capture daily, weekly and annual effects. For the annual effects we distinguish between calendar-based effects (e.g. an effect that occurs every specific calendar date, e.g. Christmas on 25 December) and effects that are driven from the meteorological cycle with a periodicity of 365.24 days. The latter contains rather smooth changes as the meteorological impact changes smoothly over the year. Further, the model contains interactions between the seasonal components, especially the daily cycle may change over the year. Next to date-based calendar effects we also include other calendar effects. Most notably holiday effects from public holiday that have a varying date, e.g. Easter Monday. The intercept of the considered model changes with all the mentioned time-varying components.
	
	The autoregressive components contain historical load data from the last hour up to the last weeks. However, we only let the most recent information to vary over time with selected time-varying components, but keep the remaining autoregressive terms constant. The autoregressive components absorb a lot of information from the past, indirectly also the information from typical external regressors like temperature. Here, we want to remark that we double-checked that the additional information of temperature in our model is negligible. In simple words: If we are at 4 pm today and want to predict the load in 1 hour for today, i.e. at 5 pm, the temperature (forecast) for 5 pm does not help a lot to improve the load forecast as the temperature information is hidden in the most recent demand observation at 4 pm, see e.g. \cite{haben2019short} for similar findings.
	
	\subsection{Augmenting structural breaks}
	Given the baseline model, we augment the time-varying intercept of the model by change-point components that allow for different types of structural breaks. 
	This is:
	\begin{itemize}
		\item[i)] a permanent change in the load level,
		\item[ii)] a permanent change in the load level for the daily profile (e.g. a load reduction for only certain hours of the day),
		\item[iii)] a permanent change in the load level for the weekly profile (e.g. a load reduction for only certain hours of the week).
	\end{itemize}
	These structural breaks are implemented using dummies for relevant time sets.	We restrict the space of possible changes to all observations after 1 March 2020 which is before the coronavirus spread widely in Europe and issued the COVID-19 crisis in Europe. We estimate the model using lasso which is tuned by the Bayesian information criterion (BIC).
	
	To analyse the results adequately, we estimate the model and then simulate from the estimated model 10000 times for the time range from 1 March 2020 onwards. This allows us to get other plausible paths of the effect. We regard the mean of the mentioned 10000 trajectories as the profile under the shutdown. We also simulate from the estimated model where we set all change-point effects to 0. This allows us to mimic a load situation without the COVID-19 shutdowns. Again, the corresponding average describes the profile that we want to compare.
	
	\section{Results}

	\subsection{Demand in Italy}
\begin{figure}[ht!]
		\includegraphics[width=1\textwidth]{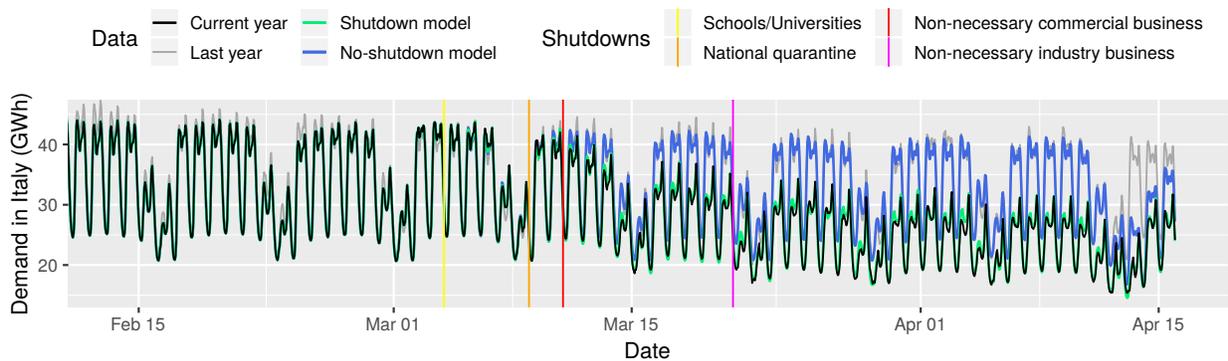}
		
		\caption{Electricity demand (GWh) in Italy during the ongoing pandemic compared to the models' and the last year's values. The vertical lines indicate shutdown dates}
		\label{fig:italy_model}
	\end{figure}

	Figure~\ref{fig:italy_model} extends Figure~\ref{fig:italy_timeline} by adding the models' and previous year's curves\footnote{Note that previous year corresponds to a shift by 52 weeks (=364 days) to represent the same weekday pattern.}. Let us note a very similar trajectory of the no-shutdown model to the last year's one. This indicates that the model is performing correctly. The only big inconsistency between these paths appears in the week starting on 13 April 2020. However, this is the week after the moveable Easter and thus a plausible public holiday effect.
	
	Moreover, we observe that the current year's electricity demand started to deviate significantly from the no-shutdown model shortly after the third shutdown and it only deepened with the fourth one. The difference between the shutdown and no-shutdown models only confirms that the undertaken measures have heavily impacted the electricity load in Italy. 
	Nevertheless, the seasonal effect is also present what is depicted by the slow decrease of the demand level of the model assuming no change-points. Hence, the shutdown effect is smaller than the naive comparison with pre-shutdown demand suggests.
	Another interesting aspect is that the structural change due to the shutdown of the non-necessary commercial business is quite smooth and requires a couple of days to settle at the corresponding load level. This suggests that after the mentioned shutdown some businesses were still running for a few days prior closing.

	\begin{wrapfigure}[13]{r}{0.5\textwidth}
		\vspace{-0.2cm}
	\includegraphics[width=1\linewidth]{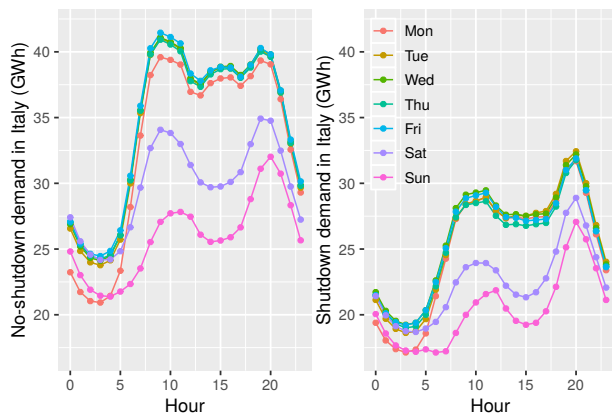}
	
	\caption{Weekly electricity demand (GWh) in Italy in the week starting on 30 March 2020 in a theoretical, no-shutdown case (left) and in the observed one (right)}
	\label{fig:italy_march}
\end{wrapfigure}

In Figure~\ref{fig:italy_march}, we present a comparison of weekly demand over hours of the day between the theoretical, no-shutdown case and the observed that includes the shutdown effects. The plots can help to understand better the change in the weekly demand pattern as they are based on the week from 30 March to 6 April 2020, i.e. during the time of a significant impact of the shutdowns. 

First, let us note the overall decrease of the load in the shutdown scenario. An interesting effect is the flattened morning peak (around 8 am - 12 am). This is most probably a result of many people working from home or not working at all and thus lesser utilization of production capacities, office building and electrified public transport etc. Interestingly, the evening peak in demand is preserved and currently it is clearly the most electricity consuming part of every day in the week. This is reasonable as because of the lockdown, more people are cooking at home or using electricity-based entertainment. Furthermore, the difference in electricity demand between Saturday and Sunday shrank heavily due to the shutdowns. 

Another interesting feature is that we see shifts of the morning load peak within the day. This is best visible on Sundays: usually at 7 am the load level would increase by about 2.5GWh ($\approx 10\%$ of the night load) from the night level. During the shutdown the increase starts later, at 7am we still remain at the night level load. A plausible explanation would be a 'getting up late'-effect. So the Italians tend to sleep longer during the lockdown period.

	\subsection{Demand in other European countries}

	\begin{figure}[p!]
	\centering
	\includegraphics[width = 1\linewidth]{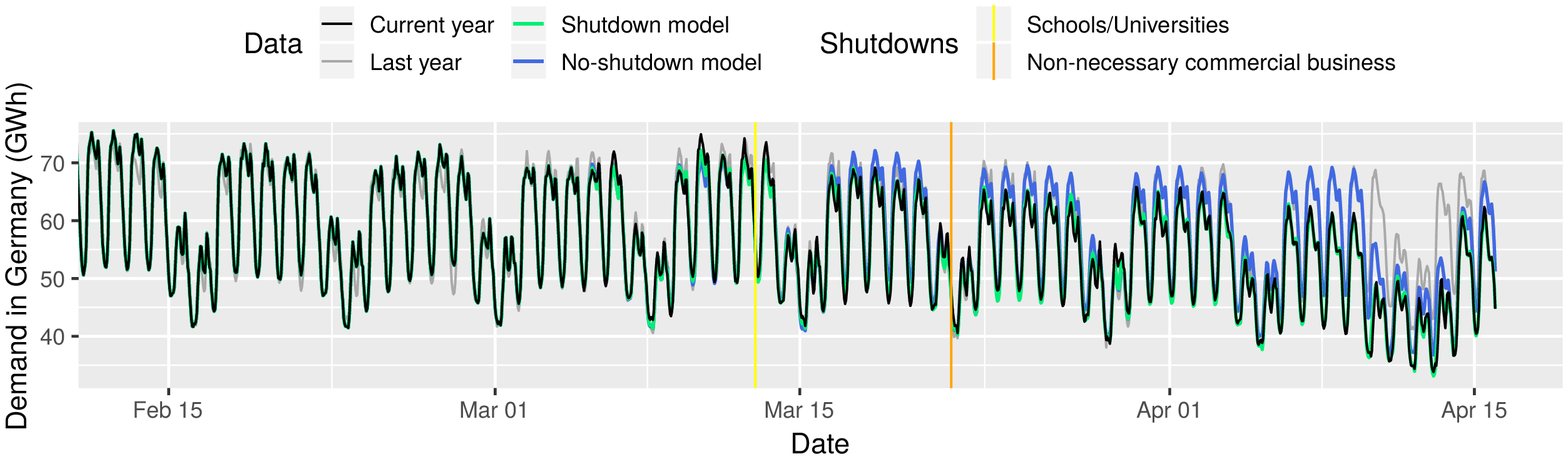}\\
	\includegraphics[width = 1\linewidth]{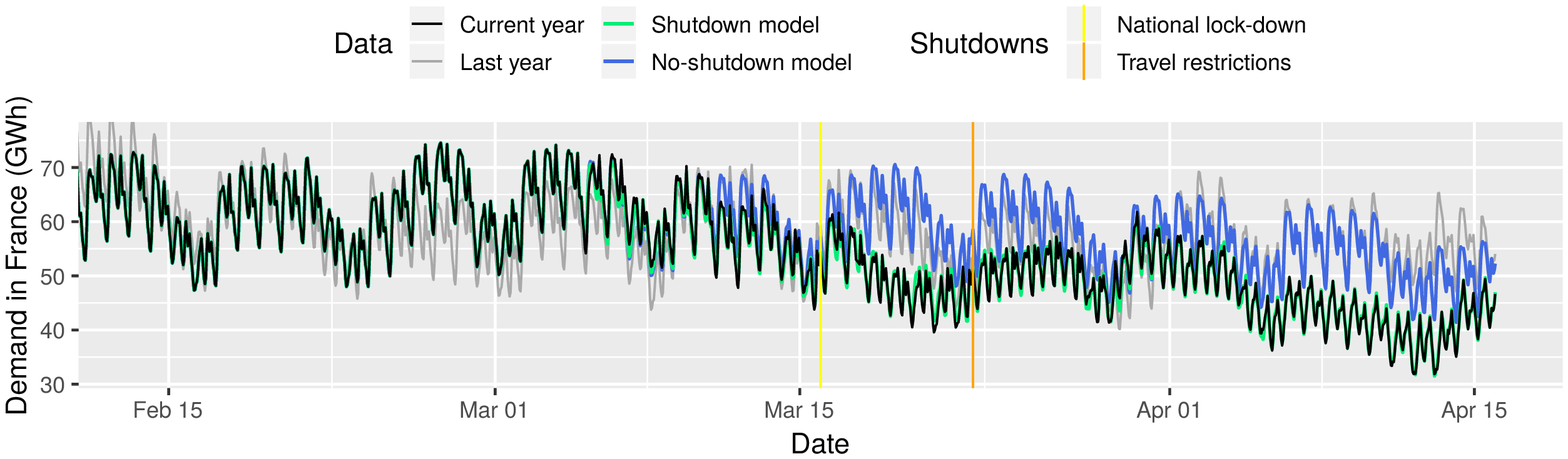}\\
	\includegraphics[width = 1\linewidth]{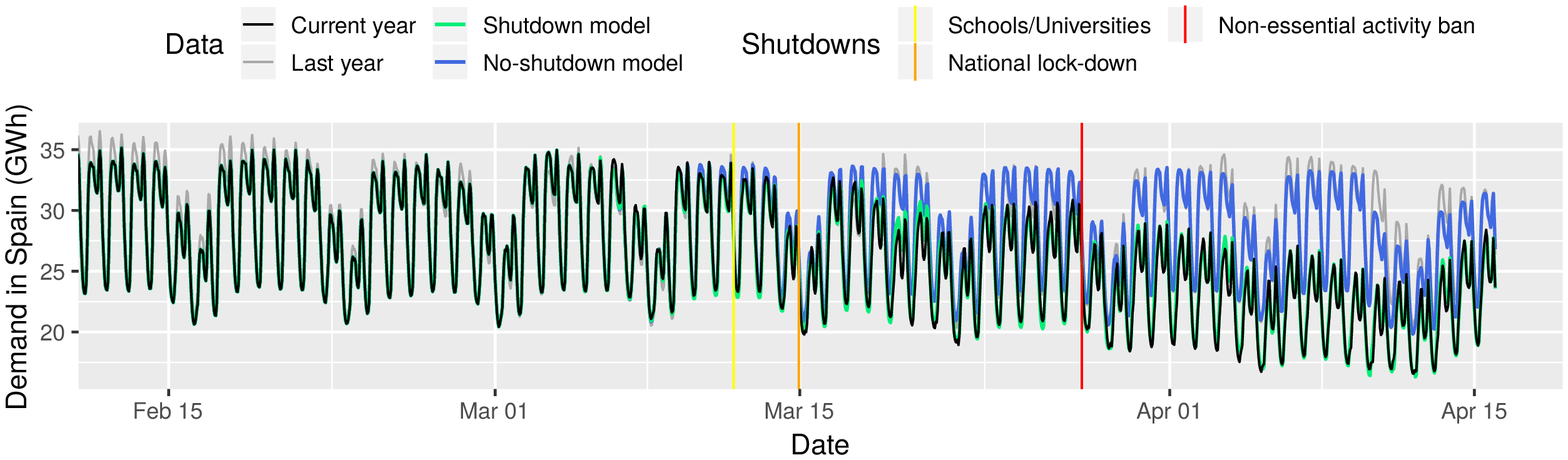}\\
	\includegraphics[width = 1\linewidth]{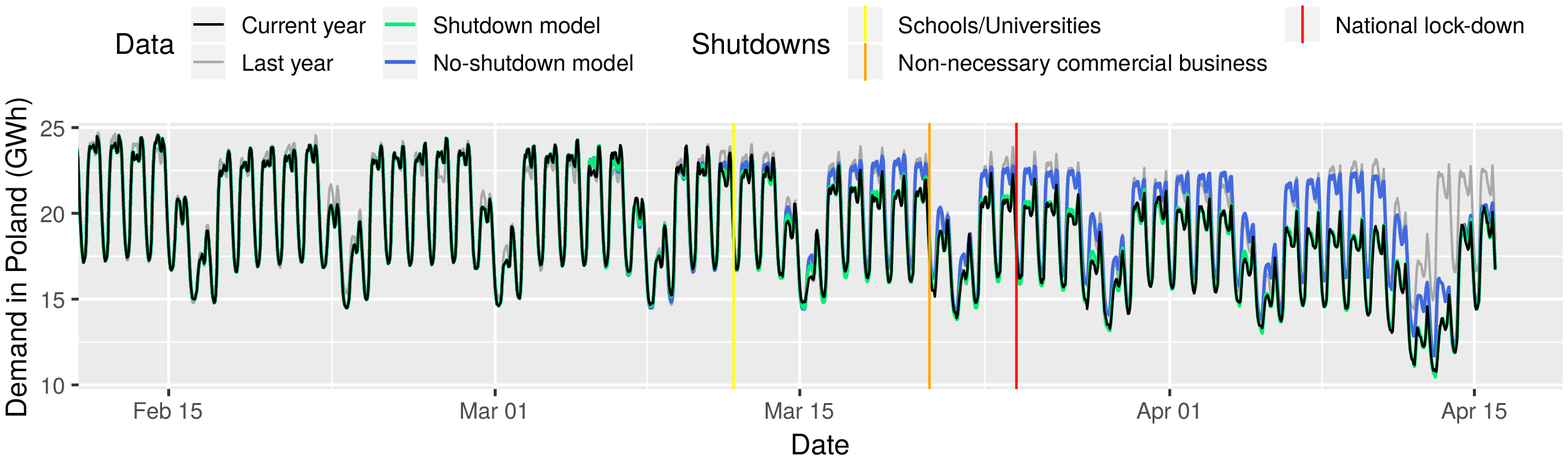}
	
	\caption{Electricity demand (GWh) in Germany, France, Spain and Poland during the ongoing pandemic compared to the models' and the last year's values. The vertical lines indicate shutdown dates}
	\label{fig:other_countries_model}
\end{figure}

	Figure~\ref{fig:other_countries_model} presents the electricity demand in the other considered countries: Germany, France, Spain and Poland. The overall pattern of the rising deviation between the shutdown and no-shutdown models is similar to the one in Italy, but respectively delayed. However, the level of the deviation differs among the states, what was already depicted in Figure~\ref{fig:motivation}. Interestingly, in France we observe an impact of the shutdowns before they went live. The reason may be that the limitations were announced accordingly earlier and the residents and companies of France may have started changing their public activity earlier, following the other countries' recommendations. Furthermore, even before the national lock-down all big events, football matches etc. were being cancelled. However, there might be interactions with the export of electricity (esp. to Italy) and temperature effects. Concerning the latter, the period from 21 March to 2 April was relatively cold in Europe, and France has a high temperature dependency in the electricity demand due to large electric heating capacities.
	
\begin{figure}[t!]
		\centering
		\includegraphics[width = 1\linewidth]{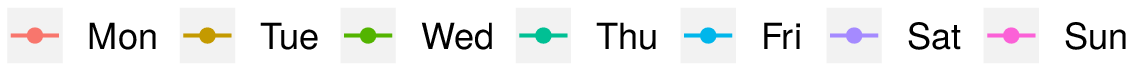}\\
		\includegraphics[width = 0.49\linewidth]{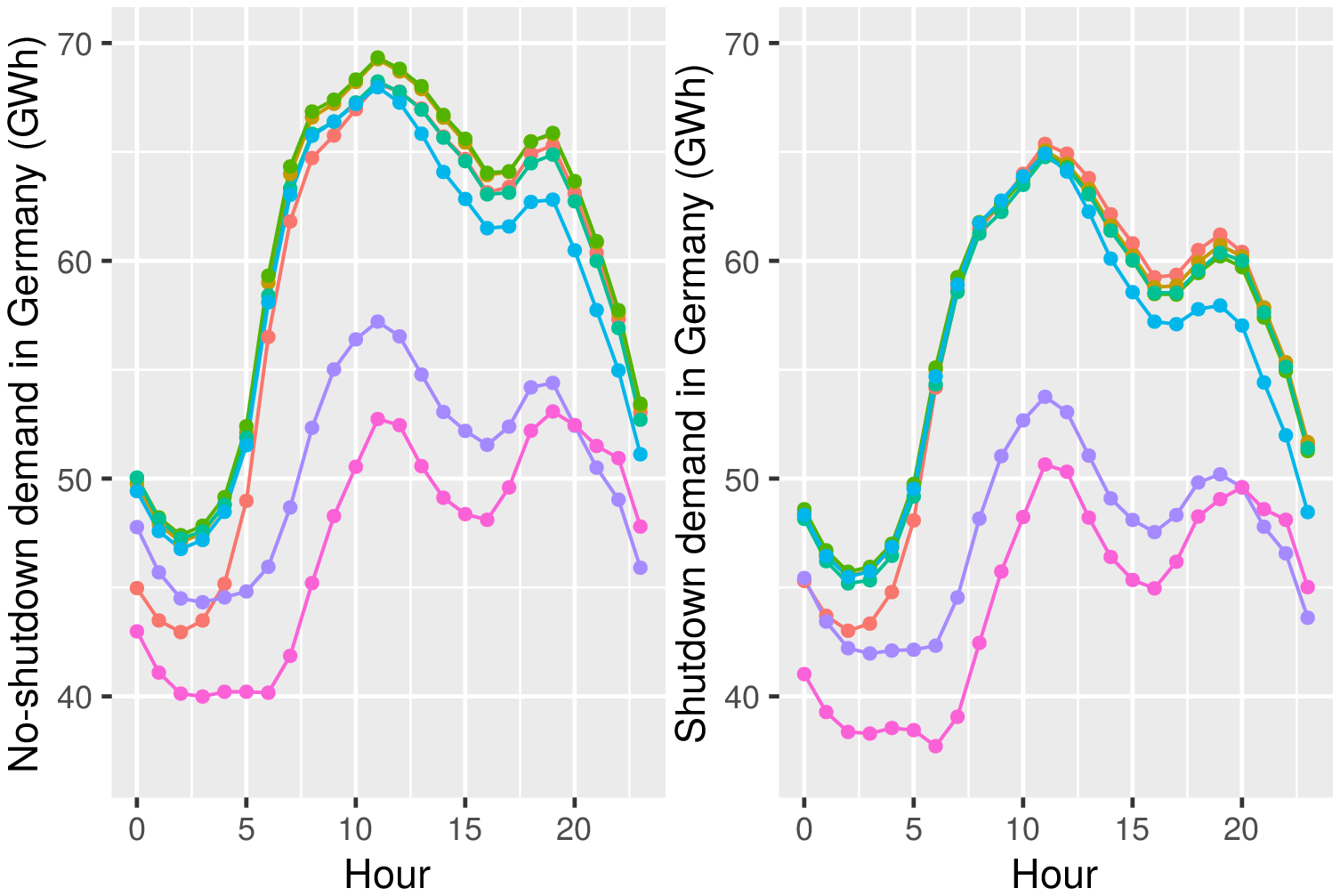}~
		\includegraphics[width = 0.49\linewidth]{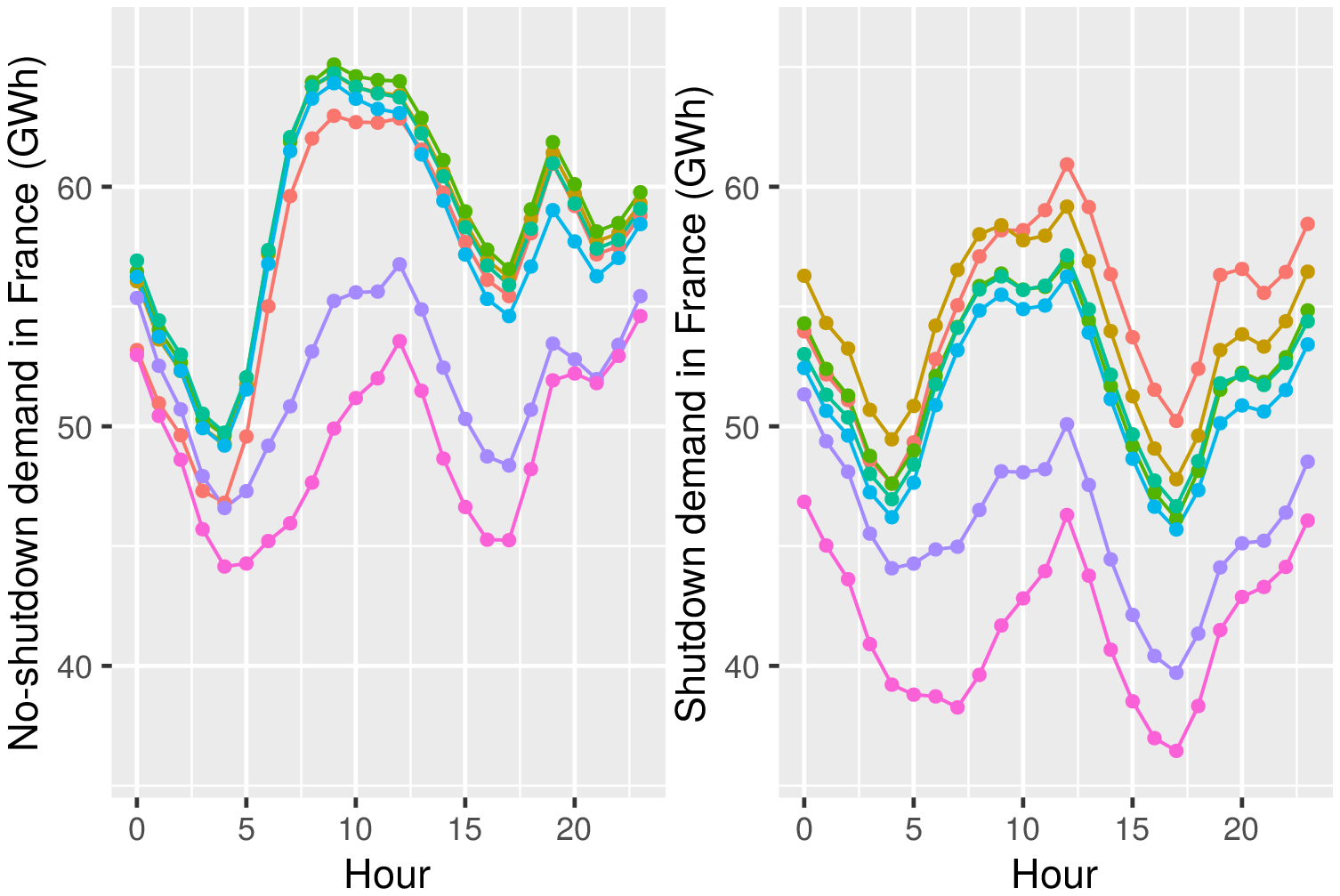}\\
		\includegraphics[width = 0.49\linewidth]{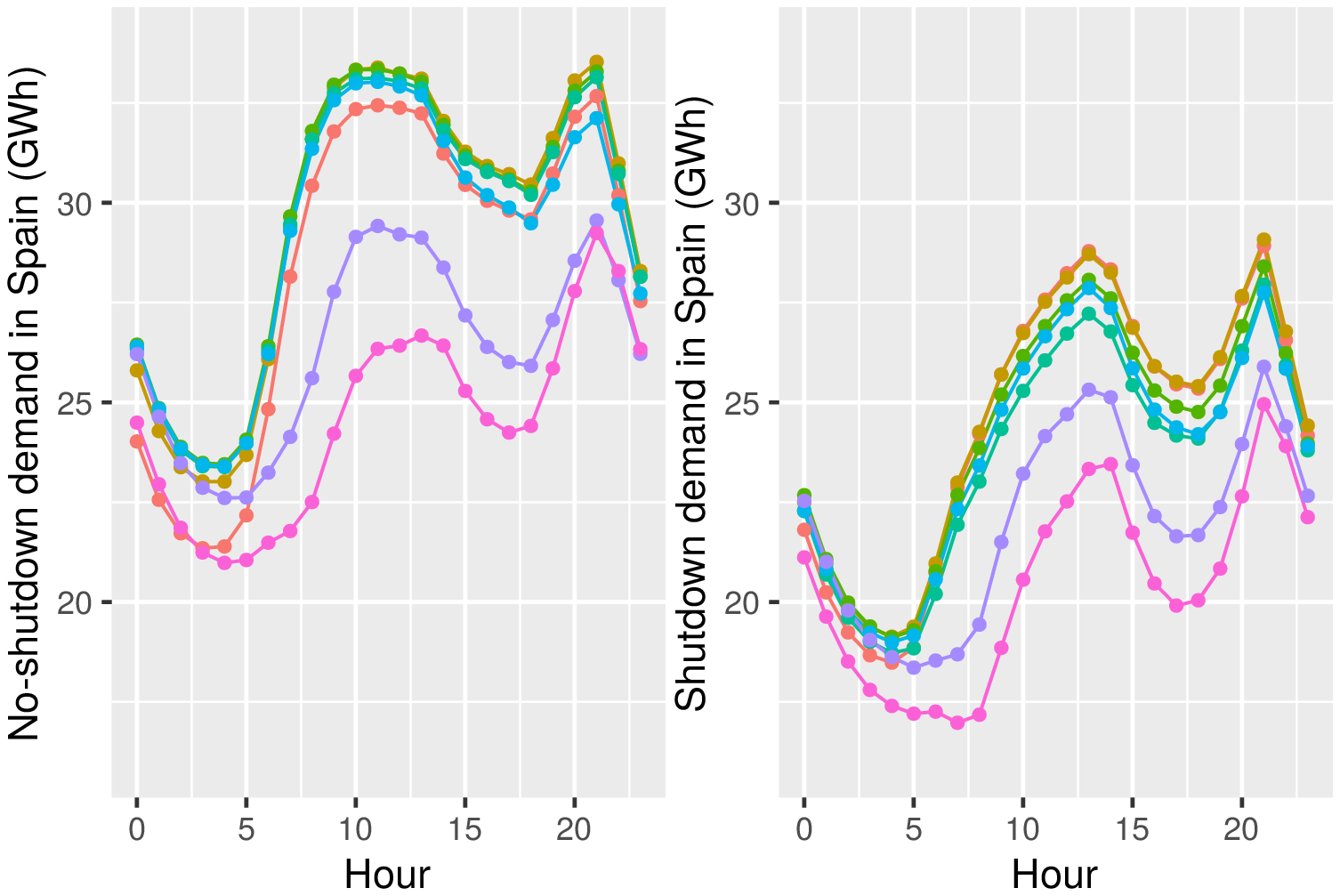}~
		\includegraphics[width = 0.49\linewidth]{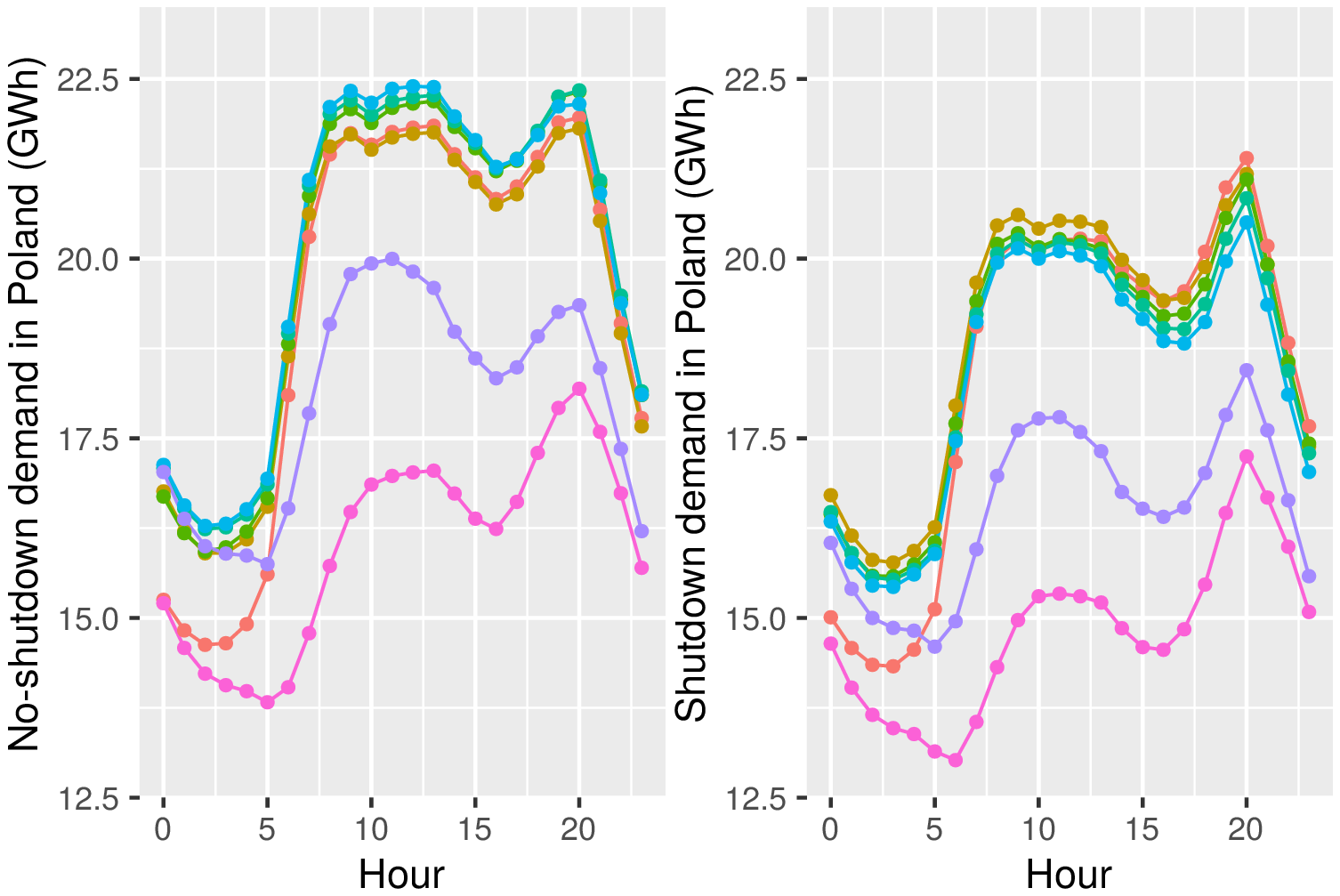}
		
		\caption{Weekly electricity demand (GWh) in Germany, France, Spain and Poland in the week starting on 30 March 2020 in a theoretical, no-shutdown case and in the observed one}
		\label{fig:other_countries_march}
	\end{figure}

	Figure~\ref{fig:other_countries_march} shows the comparison of weekly demand over hours of the day between the shutdown and no-shutdown scenarios for Germany, France, Spain and Poland. Again, the plots are based on the week from 30 March to 6 April 2020. Similarly as in the Italian case, we observe an overall demand decrease for every country. Let us note that except of the level change, the weekly demand pattern remained almost the same in Germany. On the other hand, in France, Spain and Poland the flattening of the morning peak and preserving the evening peak are present similarly as in Italy. This can be also explained by lesser activity in the morning connected to the professional life and remained or even higher activity in the evening due to entertainment. The 'getting up late'-pattern is also visible in all the  considered countries. Still, it is most distinct in the Mediterranean countries: France and Spain.
	
	\section{Conclusion}
	
	The shutdowns introduced due to the COVID-19 pandemic have impacted significantly both the level of the electricity demand in Europe and its weekly pattern. The revocation of the shutdowns and the end of the pandemic should in theory slowly turn back the electricity demand to the pre-pandemic volumes. However, in practice it may appear that the pandemic has made a permanent influence on the behavioural patterns.

	\vspace{-5mm} 
	\bibliographystyle{chicago}

	\bibliography{bibliography}	

\end{document}